\documentclass[lettersize,journal,compsoc]{IEEEtran}
\usepackage{amsmath,amsfonts}
\usepackage{algorithmic}
\usepackage{array}
\usepackage[caption=false,font=normalsize,labelfont=sf,textfont=sf]{subfig}
\usepackage{textcomp}
\usepackage{stfloats}
\usepackage{url}
\usepackage{verbatim}
\usepackage{graphicx}
\def\BibTeX{{\rm B\kern-.05em{\sc i\kern-.025em b}\kern-.08em
    T\kern-.1667em\lower.7ex\hbox{E}\kern-.125emX}}
\usepackage{balance}

\usepackage[textsize=footnotesize]{todonotes}
\presetkeys{todonotes}{inline,backgroundcolor=LightCyan}{}

\RequirePackage{todonotes}

\begin{document}
\title{Qualitative Research Methods in Software Engineering: Past, Present, and Future}
\author{Carolyn Seaman, Rashina Hoda, Robert Feldt
\IEEEcompsocitemizethanks{\IEEEcompsocthanksitem Seaman is with University of Maryland, Baltimore County, USA \protect\\
Hoda is with Monash University, Melbourne, Australia \protect\\ 
Feldt is with Chalmers University of Technology, Gothenburg, Sweden\protect\\
% note need leading \protect in front of \\ to get a newline within \thanks as
% \\ is fragile and will error, could use \hfil\break instead.
Contact E-mail: cseaman@umbc.edu}% <-this % stops an unwanted space
\thanks{Manuscript received January 2025}}

\markboth{Accepted to IEEE Transactions on Software Engineering, Feb~2025, DOI={https://doi.org/10.1109/TSE.2025.3538751}}%
{How to Use the IEEEtran \LaTeX \ Templates}

\maketitle

\begin{abstract}
The paper entitled ``Qualitative Methods in Empirical Studies of Software Engineering'' by Carolyn Seaman was published in TSE in 1999. It has been chosen as one of the most influential papers from the third decade of TSE’s 50 years history. In this retrospective, the authors discuss the evolution of the use of qualitative methods in software engineering research, the impact it's had on research and practice, and reflections on what is coming and deserves attention.
\end{abstract}

\begin{IEEEkeywords}
Qualitative Research, Software Engineering, Research Methods
\end{IEEEkeywords}

\section{Introduction}
\IEEEPARstart{T}o help celebrate TSE's 50th anniversary year, and to mark the choice of Seaman's 1999 paper \cite{seaman1999qualitative} as one of the most influential, Seaman was asked by the TSE editors to write a retrospective on the paper and its impact. Seaman was joined by Rashina Hoda and Robert Feldt to create this retrospective, using (of course) qualitative methods to elicit relevant information. The three coauthors conducted several interviews of each other about the 1999 paper, the evolution of qualitative methods in software engineering research, and current challenges and future opportunities. While these interviews were not transcribed or coded, and no methods were applied to systematically extract themes, in the following we share some of the insights we shared and discussed that pertain to the past, present, and future of qualitative research in software engineering.

\section{Background}
The 1999 paper is largely based on work done by Seaman in her Ph.D. dissertation, which was completed in 1996 (results in \cite{seaman1997empirical} and \cite{seaman1998communication}). As a Ph.D. student, she struggled to find quantitative methods and metrics to usefully describe the phenomenon she wanted to study, communication and coordination patterns among software developers. At the suggestion of her mentor, Victor Basili, she began to explore qualitative methods. She was completely self-taught in this area, as there were no qualitative methods courses or coaches available to her at the time. She relied on sources such as Eisenhardt’s paper on building theory \cite{eisenhardt1989building} and Miles and Huberman's nuts and bolts primer on qualitative coding and data analysis \cite{miles1994qualitative}. 

Qualitative research methods were new at the time for the faculty in Seaman’s CS department at the University of Maryland, including her mentor and the members of her dissertation committee. However, they were open-minded, and one committee member, Larry Votta, then a researcher at Sun Microsystems, encouraged her to write a paper to introduce qualitative methods to the empirical software engineering research community. An upcoming special issue in TSE on empirical methods in software engineering seemed like a timely venue to try. The paper was successfully published in that special issue 25 years ago.

\section{Evolution}
Initially, the paper primarily served as a single reference for researchers to start with if they wanted to explore qualitative methods. It could be cited to help justify the choice of qualitative methods, it provided some ideas for where such methods could be useful, and it gave the reader initial pointers to how they might go about applying these methods. However, this was not enough to produce an immediate groundswell of qualitative work in the field. Adoption was in fact quite slow.

The use of qualitative methods did not increase significantly until about 10 years later. Around 2008 and 2009, we saw the first editions of ICSE co-located events such as the Cooperative and Human Aspects of Software Engineering (CHASE) and Socio-Technical Congruence, as well as others. Such events highlighted the importance of studying people who play a role in software development, not just the artifacts they produce. A new interest in studying human factors led to openness to qualitative methods, and at the same time, qualitative methods allowed for meaningful study of human factors. An increase in studies of agile software development, especially with industry collaborators, helped fuel this cycle. Also, publications such as Easterbrook et al.’s \cite{easterbrook2008selecting} chapter on selecting research methods included significant content on qualitative methods, which was a useful and clever way to get such methods into the mainstream.

By the late 2010’s, there was a noticeable positive shift in the ability to publish qualitative work in top software engineering venues and for some to receive awards, for example, Hoda’s qualitative work \cite{hoda2017becoming} received a distinguished paper awards at ICSE 2017. Sometimes this openness was led, or at least influenced, by a qualitative ally in a position to make a difference, such as an Associate Editor or a Program Chair. Several such folks, who recognized the value and potential of studying important topics qualitatively, helped to carefully choose reviewers, and also to judiciously weigh feedback from reviewers with little background, but lots of suspicion, toward qualitative methods. Storey’s ICSE keynote in 2019 
\cite{storey2019publish} was a watershed moment, as one of the first (if not the first) times that qualitative methods received significant attention on the ICSE main stage.

The 1999 paper we are reflecting on did not create an immediate paradigm shift in software engineering research. Seaman spent many years after its publication arguing for, explaining, justifying, defending, teaching, and promoting qualitative methods in all kinds of gatherings of researchers. But this campaign would not have been possible without the publication of the TSE paper in 1999. Over time, the job became easier as others joined, but it took some time. However, the original paper served, and continues to serve, as a starting point, not only for the history of the adoption of qualitative research in software engineering, but also for researchers who want to learn more. 

\section{Impact}
The greater and better use of qualitative methods has enabled many lines of knowledge creation that would not otherwise have been possible. As mentioned earlier, there is a cyclic chicken-and-egg sort of dynamic between the desire and need to study human aspects of our discipline, and the acceptance of and need for new methods to carry out those investigations. The increased acceptance of qualitative methods by the community has been both facilitated and spurred by the increased interest in human aspects.

Probably nowhere is this phenomenon more pronounced than in the growth and success of the CHASE community. CHASE (Cooperative and Human Aspects of Software Engineering) started in 2008 as an ICSE co-located workshop, and then grew to a full conference, while still co-located with ICSE. CHASE has matured into a high quality home for human-focused, qualitative studies, and has provided a forum for like-minded researchers to discuss this type of work. Researchers have been pleasantly surprised to find their `tribe' at CHASE, i.e. other people who were exploring qualitative methods and similar questions. CHASE provides a safe space for presenting qualitative work and experimenting with new methods. Its mentoring and round-table discussion sessions provide much needed forums for the community to discuss methodological issues, challenges, and workarounds.  

With the growth in importance of agile methods in practice, we've seen a substantial rise in recent years in studies of agile practices \cite{hoda2018rise}. With a focus on human, social, and socio-technical aspects, researchers studying agile topics contributed heavily to introducing, providing examples of, and using various qualitative approaches, often borrowed from social sciences. The XP (agile software development) conference community has been at the forefront of this progress, providing a quality venue for both research and industry perspectives.

Distributed software development is another area that naturally fostered early qualitative research work~\cite{jimenez2009challenges}.
The emphasis was on understanding the dynamics of how teams work together when face to face interaction is not possible, a reaction to the increased prevalence of outsourcing arrangements and other types of distributed situations. However, this work took on new meaning, and new impact, when distributed work necessarily became the norm during the pandemic~\cite{ralph2020pandemic}. It can be argued that the software industry was better prepared, and faster, to pivot to remote work because of the understanding of distributed teams provided by earlier qualitative work on distributed global development.

While Seaman’s 1999 TSE paper continues to be heavily cited, there are far more and better resources available now to qualitative researchers in software engineering, written by and for software engineering researchers. We will not attempt to provide an exhaustive list, but we’d like to highlight a few that occupy special and unique roles. Mentioned earlier is Easterbrook et al.’s \cite{easterbrook2008selecting} book chapter on \textit{selecting research methods}, which puts qualitative methods in the context of empirical design. \textit{Case study research} is tailored and explained for application in software engineering research by Runeson et al. in their  2009 Empirical Software Engineering (EMSE) article \cite{runeson2009guidelines} and in detail in their 2012 book \cite{runeson2012case}. In their 2016 TSE article, Sharp et al. explained \textit{ethnography} as a valuable method for software engineering researchers \cite{sharp2016role}. Grounded Theory is a qualitative approach that has been used and misused many times by software engineering researchers over the last 25 years, and is usefully adapted by Hoda in her 2022 TSE article introducing \textit{Socio-Technical Grounded Theory (STGT)} \cite{hoda2022stgt}. In her 2024 book \cite{hoda2024qualitative}, she presents the foundations of qualitative research and details of STGT with examples and tips. Lenberg et al.~\cite{lenberg2024qualitative} reviewed the use of qualitative methods in software engineering and argued for use of a broader set of methods such as quality checklists from the sociology and psychology literature, specifically narrative and discourse analysis, and the COREQ checklist~\cite{tong2007consolidated}. They also pushed for acknowledgement and consideration of reflexivity. These contributions made over the last 25 years are summarized in Table \ref{tab:QualMethodsGuides}. Several special issues have served to bring together a host of qualitative methods papers in Information and Software Technology in 2007  \cite{dittrich2007editorial} and in EMSE in 2011 \cite{dybaa2011qualitative}. Additionally, the awareness and understanding of qualitative methods have been further supported through several panels, workshops, tutorials, and technical briefings held at  mainstream software engineering conferences, providing a lively environment for relevant discourse and discussions.

It is still useful to look to other disciplines, many of which are more advanced in the quality and variety of qualitative work, for guidance. Of note is the Equator Network from health sciences, which provides guidance on reporting a wide variety of types of qualitative studies. In addition to their COREQ checklist for reporting on interviews and focus groups their more general SRQR checklist synthesizes guidelines for how to report qualitative research, in general~\cite{o2014standards}. However, not all qualitative researchers agree that checklists are the way forward; a recent discussion of the pros and cons, as well as a consolidated guideline, can be found in Twining et al.~\cite{twining2017some}.

\begin{table*}[]
    \centering
    \caption{A selection of qualitative research methods guides for software engineering published as journal articles, book, and book chapters (1999--2024)}
    \begin{tabular}{>{\raggedright\arraybackslash}p{11.3cm}>{\raggedright\arraybackslash}p{3.3cm}>{\raggedright\arraybackslash}p{1.8cm}}
    \hline\noalign{\smallskip}
    \textbf{Coverage of Qualitative Research Concepts, Methods, Techniques} & \textbf{Reference} & \textbf{Nature of Contribution}\\
    \hline\noalign{\smallskip}
    \noalign{\smallskip}\noalign{\smallskip}
    Overview of qualitative methods and techniques relevant for software engineering, including participant observation, interviewing, coding, generation and confirmation of theory, and experimental design. &  Seaman, 1999 \cite{seaman1999qualitative} & TSE Journal \\
    \noalign{\smallskip}
    Overview of empirical methods relevant for software engineering, including research philosophy, role of theory building, and overview of qualitative methods such as case studies, survey research, ethnography, and action research.  &  Easterbrook et al., 2008 \cite{easterbrook2008selecting} & Book chapter\\
    \noalign{\smallskip}
    Case study research method for use in software engineering, including characteristics of research methodology, case study research process and definitions, ethical considerations, data collection and analysis, reporting, reading, and reviewing case studies. & Runeson et al., 2009 \cite{runeson2009guidelines} & EMSE Journal\\
    \noalign{\smallskip}
    Case study research method for use in software engineering, including case study origins, design, data collection, data analysis and interpretation, and examples of case studies. & Runeson et al., 2012 \cite{runeson2012case} & Book\\
    \noalign{\smallskip}
    Ethnography research method for use in software engineering, including origins from sociology and other disciplines, main features of ethnographic research, dimensions of ethnographic studies, role of ethnographic studies in software engineering.  & Sharp et al., 2016 \cite{sharp2016role} & TSE Journal \\
    \noalign{\smallskip}
     Introduction to socio-technical grounded theory (STGT) research method for use in software engineering, and an overview of traditional grounded theory methods and their limitations. & Hoda, 2022 \cite{hoda2022stgt} & TSE Journal\\
    \noalign{\smallskip}
    Foundations of research design, philosophy, and literature review; basics of qualitative research: data collection techniques, preparation, filtering; socio-technical grounded theory for qualitative data analysis; theory and theory development; evaluation guidelines; future of qualitative research with LLMs. & Hoda, 2024 \cite{hoda2024qualitative} & Book\\
    \noalign{\smallskip}
    Review of qualitative methods in software engineering: grounded theory and thematic analysis, dominating on primarily interview data; use of sociological/psychological methods and quality criteria in software engineering, e.g. reflexivity, COREQ checklist, and  narrative\slash discourse analysis. & Lenberg et al, 2024 \cite{lenberg2024qualitative} & JSEP Journal\\
    \noalign{\smallskip}\noalign{\smallskip}
    \hline\noalign{\smallskip}
    \end{tabular}
    \label{tab:QualMethodsGuides}
\end{table*}

\section{Current Issues and Future Directions}

\begin{table*}[ht]
\centering
\caption{Recommendations for Advancing Qualitative and Mixed-Methods Research in Software Engineering}
\begin{tabular}{|p{0.05\textwidth}|p{0.85\textwidth}|}
\hline
\textbf{ID} & \textbf{Recommendation} \\ \hline
1 & \textbf{Adopt a Variety of Methods:} Utilize a variety of qualitative, quantitative, and mixed methods approaches to study phenomena from diverse perspectives. This allows for richer, more comprehensive insights and addresses complex research questions. \textit{As editors and reviewers, recognize and value the use of many different methods, focusing on their fit to the goals of the research.}  \\ \hline
2 & \textbf{Learn from Other Disciplines:} Continue to draw inspiration from sociology, psychology, anthropology, and other fields to adopt and adapt established methods and frameworks that address socio-technical challenges. \\ \hline
3 & \textbf{Engage with Real-World Stakeholders:} Collaborate with industry, developers, and end-users to address practical challenges, where possible. Actively include their perspectives when proposing technical solutions or writing papers to ensure that research aligns with real needs and has practical and positive societal impact.  \\ \hline
4 & \textbf{Innovate Methodologically:} Explore and adapt qualitative methods to the unique socio-technical context of software engineering while maintaining methodological rigor. \textit{As editors and reviewers, support and reward innovation without raising unnecessary obstacles to new methodological approaches. Actively promote venues, awards, and opportunities that encourage methodological experimentation.} \\ \hline
5 & \textbf{Encourage Creativity:} Push the boundaries of creativity in research design and reporting. Balance innovation with rigor to uncover new perspectives and insights, while fostering an open dialogue about the acceptable bounds of creativity. \textit{As editors and reviewers, strive to appreciate new and creative ways to achieve research objectives, stretching the boundaries of what is the norm.} \\ \hline
6 & \textbf{Highlight Ethical Considerations:} Ensure studies meet ethical standards, protecting participant anonymity and responsibly handling sensitive data. \textit{As reviewers and editors, embed ethical requirements into submission and review processes. As a community, ensure that ethical standards make sense for all types of research.}  \\ \hline
7 & \textbf{Advance Reflexivity:} Explicitly consider and report philosophical stances, biases, and assumptions during research design and reporting. This enhances transparency, credibility, and trust in qualitative work. \\ \hline
8 & \textbf{Focus on Emergence:} Embrace emergent results that challenge prior theories and assumptions. Inductive approaches can reveal novel insights, even if they do not conform tightly to initial research questions, or even emerge without initial questions at all. \\ \hline
9 & \textbf{Pursue Longitudinal and Large-Scale Studies:} Seek and take advantage of opportunities to conduct studies that examine phenomena over time or at organizational and ecosystem levels. These approaches provide deeper understanding of dynamic and systemic issues and evolution of practice over time.  \\ \hline
10 & \textbf{Leverage New Tools Responsibly:} Experiment with AI-enabled tools, such as Large Language Models (LLMs), for qualitative data collection and analysis. Ensure their integration is warranted, ethical, thoughtful, and enhances—not undermines—the validity and value of the research.  \\ \hline
11 & \textbf{Set and Promote Tailored Guidelines:} Work towards tailored methodological and reporting guidelines for qualitative and mixed methods research in software engineering. \textit{As editors and reviewers, encourage adherence to such guidelines to ensure consistency and clarity. As a community, develop guidelines that expand, not restrict, the methodological choices available to researchers.} \\ \hline
12 & \textbf{Mentor and Build Community:} Create opportunities for collaboration and knowledge-sharing, such as workshops, tutorials, panels, roundtables, and special issues, to support qualitative researchers and promote methodological diversity.  \\ \hline
\end{tabular}
\label{tab:recommendations}
\end{table*}

As we reflect on the current level of maturity of qualitative work in software engineering, it is gratifying to see how far it has come. It is also gratifying to see the pace of innovation in this area. Many qualitative researchers in software engineering continue to innovate and sample from the wide variety of methods used and described in other disciplines. This is not easy to do, as there is always a risk in being the first to apply a novel (to software engineering) method. There is the risk that it will not be accepted, or will be criticized too heavily, or not understood. There is also the risk of not applying the method well and appropriately, made even worse if the work is widely accepted and the method (badly applied) is widely adopted. Thus, researchers willing to step out and try new things are to be applauded and supported. More published examples of a wider variety of approaches can only benefit the field.

Most qualitative methods in software engineering originate from the social sciences, often adapted on the fly to fit the unique socio-technical context, with varying success \cite{hoda2022stgt}. Experienced researchers have developed tailored method guidelines for software engineering (see Table \ref{tab:QualMethodsGuides}), but these are frequently met with resistance. Methodologists face repeated demands to justify adaptations specific to software engineering, even when such innovations bridge critical methodological gaps and enable more robust research.

A new method is not a goal in itself, nor should its value be assumed simply because it is novel. Methods are developed for specific purposes and contexts, and their relevance depends on whether they address the challenges of the situation at hand. While not every new method will be universally applicable, many are well designed to fulfill a research need and can provide valuable insights when thoughtfully applied. This complexity underscores the fact that methodology choices are rarely black and white.

Once established, software engineering-specific guidelines could serve as the primary reference for researchers. However, reviewers sometimes insist on citing original social science sources instead, adding unnecessary burdens to researchers exploring new methodological territory. This reflects a broader issue of entrenched ``methodology camps,'' where researchers rigidly defend favored approaches and resist unfamiliar methods, often leading to unproductive debates over which methodologies are superior. We are not only talking about quantitative researchers reviewing qualitative research here; it's not uncommon for qualitative researchers to implicitly or explicitly cling to a chosen methodological ``camp''.

We advocate for a balanced and open perspective, recognizing that no single methodology suits all inquiries. The value of carefully crafting software engineering-specific guidelines to adapt and apply methods from other disciplines cannot be overstated---they are essential for advancing the field while maintaining methodological rigor and relevance. Even if a methodology needs little adaptation when applied in software engineering, this is important to explain and discuss.

It’s also tempting to envision the many, many ways that qualitative work could advance and become more impactful to software development in practice. Better use of mixed methods, for example, could allow us to generate richer, better validated understanding of phenomena. Forthcoming guiding principles from Storey et al. \cite{storey2025guidelines} on mixed methods treat qualitative methods in the context of the spectrum of research methods. Longitudinal studies are extremely difficult, but so important as most truly interesting phenomena happen over time, and snapshots are limited in facilitating understanding. Advances in, and acceptance of, analysis of video data is needed. This would facilitate investigations in contexts that are otherwise very hard to access. Raising the scope of studies from the individual or team level, so that we begin to study whole organizations and ecosystems, could be possible using qualitative methods and would yield new knowledge at a different level~\cite{lenberg2015behavioral}. Qualitative researchers in software engineering also have an opportunity to work very closely with industry, to listen (because we're good at that) to the concerns and burning issues in practice, and to use methods that practitioners intuitively trust and value.

It is also time to pay more careful consideration to higher-level concerns related to research design, and the use of qualitative methods in particular. One is reflexivity, i.e. the explicit consideration of the philosophical stance and biases of the researcher~\cite{lenberg2024qualitative}. Taking emergence seriously is another. Allowing for truly emergent results (i.e. inductive analysis without preexisting research questions or hypotheses) allows the researcher, and the audience, to hear data and results that challenge our assumptions, our implied knowledge structures, and our prior theoretical positions. This is also true of quantitative work, so the qualitative research community might have an opportunity to help all researchers use emergence well to facilitate knowledge generation in many areas. Acceptance of creativity in research philosophy and in the analysis and reporting of qualitative work is another issue that warrants a conversation. Creativity risks eroding rigor in some cases, but can also reveal insights and perspectives that lie outside our familiar templates and checklists. The increased prevalence of open science standards have, in some cases, taken into account the unique needs of qualitative research, but more needs to be done. Ethical considerations in qualitative research are paramount, particularly in ensuring informed consent, protecting participant anonymity, and minimizing potential harm when analyzing and presenting sensitive data. Researchers must also remain vigilant against biases and power imbalances, fostering trust and respecting the rights of participants throughout the study. Some guidance is available~\cite{strandberg2019ethical} but broader consideration of ethics is needed in software engineering research, be it qualitative or not.

Qualitative research in software engineering, like any other endeavor in 2025, must address and explore the role of artificial intelligence. At this point in time, the AI-enabled technology that seems most promising, and possibly most scary, is the use of Large Language Models (LLMs) to aid in the generation, coding, analysis, and sense-making of qualitative data \cite{bano2024large, hoda2024qualitative}. This should not be dismissed as a lazy shortcut and a threat to validity. Instead, we should critically examine how the purposeful use of LLMs could enhance creativity, emulate (and thus reveal) our biases, and of course make us more productive. As empiricists, there are rich opportunities to test and explore the boundaries of what is possible and wise. 

\section{Conclusion} 
It is gratifying to see, now in 2025, how much good qualitative work has been published and what new knowledge has resulted in the last quarter century. But what's truly exciting is the potential of the qualitative research approach, and the possibilities going forward. The software engineering research community has no shortage of thorny, practical, real problems to work on, and no shortage of opportunities to impact practice. Therefore, there is no reason not to continue to expand our methodological repertoire even further.

This paper reflects on the history and the current status of the use of qualitative methods in software engineering. Looking to the future, Table~\ref{tab:recommendations} outlines a consolidated set of actionable recommendations, aimed at researchers, reviewers, editors, organizers, and the software engineering research community at large. We intend these \textit{wishes} and \textit{dreams} for the future to serve as a source of inspiration and, hopefully, a call to action. They emphasize the potential of diverse methodologies, engagement with real-world stakeholders, cross-disciplinary learning, and innovation in qualitative research practices. Additionally, they highlight the importance of reflexivity, ethical rigor, and creativity, while encouraging openness, collaboration, and recognition within the research community. By collectively considering these ideas, we hope to inspire further progress in using qualitative methods for understanding the complex socio-technical phenomena that define our field and enhance the impact of our work.

In conclusion, we'd like to leave our readers with some questions, based on our discussion in this retrospective, to consider and debate going forward. 

\begin{itemize}
    \item What areas of software engineering practice have been most impacted by qualitative research? Which areas could benefit from more qualitative work?
    \item How can we balance the need for methodological innovation with the desire to apply new methods appropriately and correctly? Can we support innovators while avoiding the emulation of `bad` examples in the literature?
    \item Can methods for research involving modern sources of data (e.g., video, immersive, AI generated) be adapted from other disciplines for software engineering research? 
    \item How can qualitative researchers better leverage the natural advantages they have in working closely with industry?
    \item What changes need to be made to open science standards to fully embrace the needs of qualitative researchers?
    \item Should there be a distinction between \textit{hard earned} data (e.g., through interviews, observations) versus data that is generated by LLMs, in terms of quality, credibility, authenticity, and acceptability for different research scenarios? 
    \item What are the most useful ways to use LLMs and other AI-enabled technologies in the research process? What, if any, bounds should be placed on acceptable use?
\end{itemize}

We look forward to the conversation! 

\bibliographystyle{IEEEtranS}
\bibliography{editorial.bib}

\section*{Author Biographies}
\vspace{-4.7cm}
\begin{IEEEbiography}[{\includegraphics[scale=0.09]{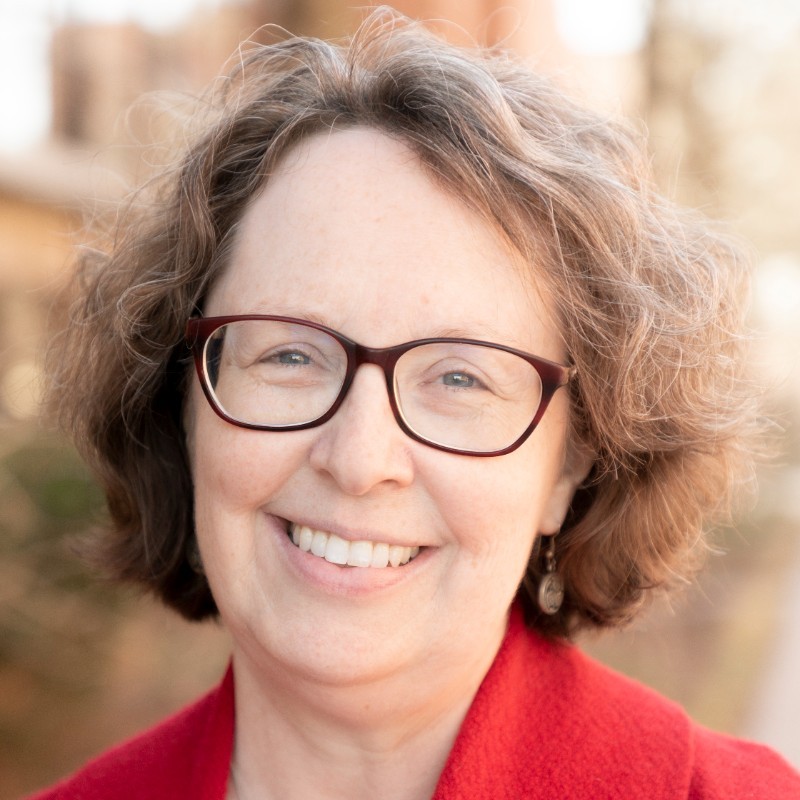}}]{Carolyn Seaman} is a Professor of Information Systems at the University of Maryland Baltimore
County (UMBC). She is also the Director of the Center for Women in Technology, also at UMBC.
Her research consists mainly of empirical studies of software engineering, with particular emphases on
maintenance, organizational structure, communication, measurement, and technical debt. She also
investigates research methodology in software engineering, as well as computing pedagogy. She
holds a PhD in Computer Science from the University of Maryland, College Park, a MS from Georgia
Tech, and a BA from the College of Wooster (Ohio).
\end{IEEEbiography}
\vspace{-4.7cm}
\begin{IEEEbiography}[{\includegraphics[scale=0.14]{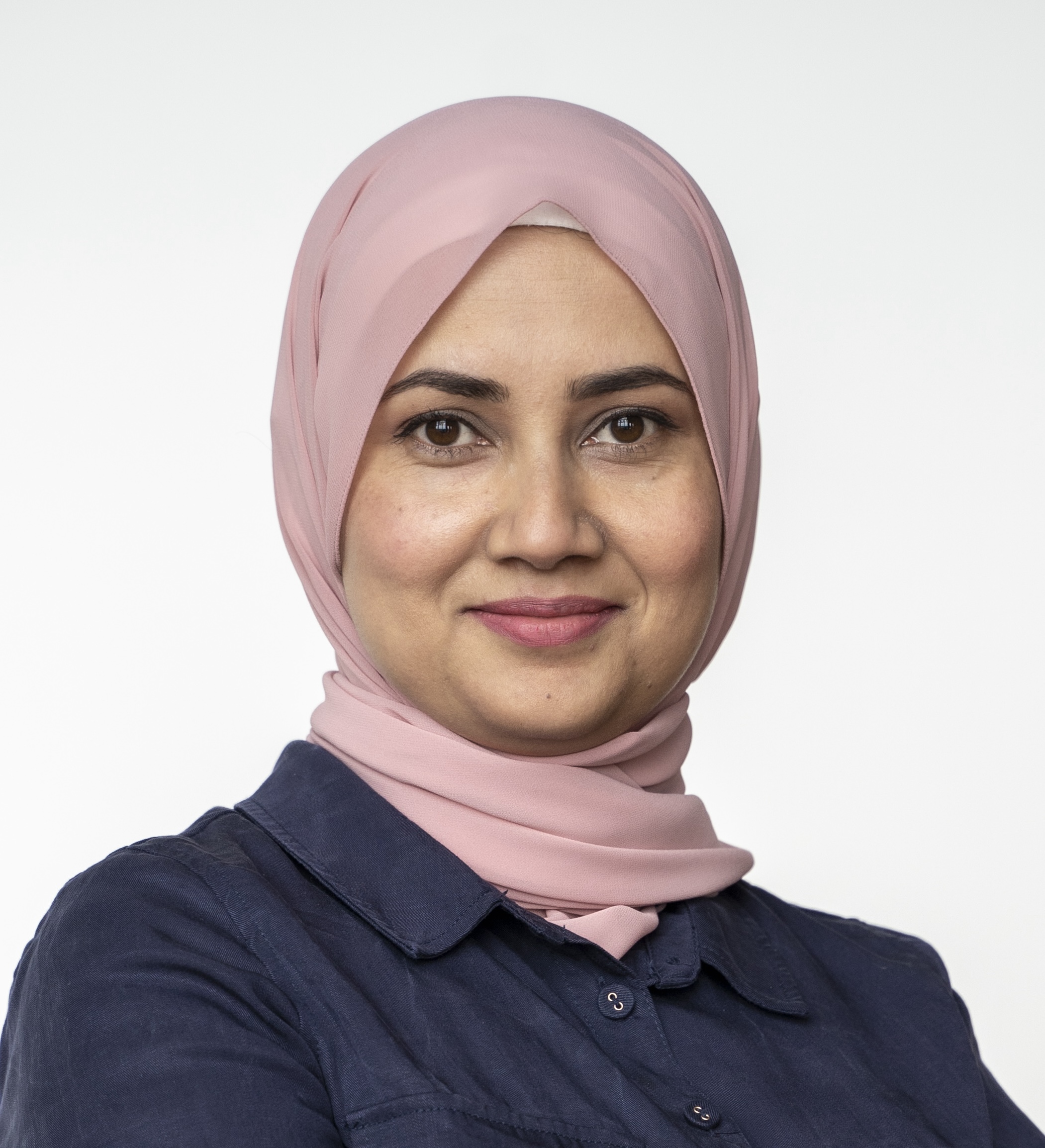}}]{Rashina Hoda} is a Professor of Software Engineering at Monash University, Australia. Her research focuses on the human and socio-technical aspects of Software Engineering, Artificial Intelligence, and Digital Health, using qualitative and mixed methods research approaches. She was named the 2025 Top Researcher in Software Systems in Australia by The Australian. In her 2024 Springer book ``Qualitative Research with Socio-Technical Grounded Theory'', she presents a modern socio-technical variation to the Grounded Theory methods for Software Engineering and other fields. Rashina serves as an Associate Editor of the IEEE Transactions on Software Engineering and the General Chair for CHASE 2025. More about her on www.rashina.com
\end{IEEEbiography}
\vspace{-4.7cm}
\begin{IEEEbiography}[{\includegraphics[scale=0.22]{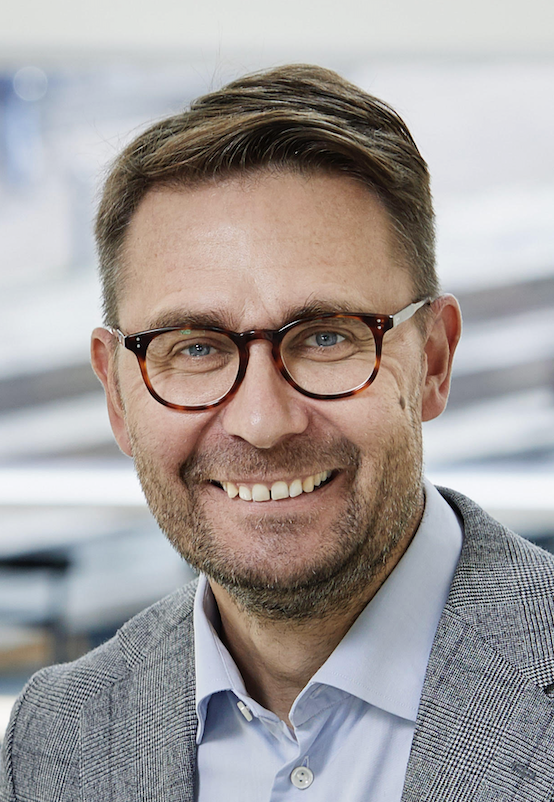}}]{Robert Feldt} is a Professor of Software Engineering at Chalmers University of Technology in Gothenburg, Sweden. His research focuses on software testing, quality, and human-centered software engineering, with passionate interests in human factors, automation, applied machine learning (ML), and a broad range of research methodologies. He collaborates with companies across Sweden, Europe, and Asia as an innovation and AI strategist while also leading foundational research. He holds a PhD in Computer Science and Engineering from Chalmers University of Technology. He serves as Co-Editor-in-Chief of Springer’s Empirical Software Engineering journal since 2017 and is on the editorial boards of STVR and SQJ.
\end{IEEEbiography}

\end{document}